   \documentclass[showpacs,prb,superscriptaddress,twocolumn]{revtex4}
\usepackage [cp1251]{inputenc}
\usepackage{bm}
\usepackage{amsmath}
\usepackage{amssymb}
\usepackage{graphicx}
\newcommand{\A}{\mathcal{A}}
\newcommand{\B}{\mathcal{B}}
\newcommand{\F}{\mathcal{F}}
\newcommand{\TM}{\mathcal{TM}}
\newcommand{\MT}{\widetilde{\mathcal{TM}}}
\newcommand{\arctanh}{\mathop\mathrm {arctanh}\nolimits}
\newcommand{\e}{{\rm e}}
\renewcommand{\i}{{\rm i}}

\begin{document}

\title{Resonant Photonic Quasicrystalline and Aperiodic Structures}
\author{A.N.~Poddubny}\email{poddubny@coherent.ioffe.ru}
\affiliation{A.F.~Ioffe Physico-Technical Institute, 194021
St.~Petersburg, Russia}
\author{L.~Pilozzi}
\affiliation{Istituto dei Sistemi Complessi, CNR, C. P. 10,
Monterotondo Stazione, Rome I-00016, Italy}
\author{M.M.~Voronov}
\affiliation{A.F.~Ioffe Physico-Technical Institute, 194021
St.~Petersburg, Russia}
\author {E.L.~Ivchenko}
\affiliation{A.F.~Ioffe Physico-Technical Institute, 194021
St.~Petersburg, Russia}

\begin{abstract}
We have theoretically studied propagation of exciton-polaritons in
deterministic aperiodic multiple-quantum-well structures,
particularly, in the Fibonacci and Thue-Morse chains. The
attention is concentrated on the structures tuned to the resonant
Bragg condition with two-dimensional quantum-well exciton.  The superradiant or
photonic-quasicrystal regimes are
realized in these structures depending on the number of the
wells. The developed theory based on the two-wave
approximation allows one to describe analytically the exact transfer-matrix
computations for transmittance and reflectance spectra in the
whole frequency range except for a narrow region near the
exciton resonance. In this region
the optical spectra and the exciton-polariton dispersion
demonstrate scaling invariance and self-similarity which can be
interpreted in terms of the ``band-edge'' cycle of the trace map,
in the case of Fibonacci structures, and in terms of zero
reflection frequencies, in the case of Thue-Morse structures.
\end{abstract}
\maketitle

\section{Introduction}

Quasicrystalline and other deterministic aperiodic  structures are
one of the modern fields in photonics
research.\cite{Merlin1985,Kokmoto1986,Ledermann2006,Matsui2007}
Due to a long-range order such structures can form wide band
gaps in energy spectra as in periodic photonic
crystals\cite{Yablonovich1987,John1987} and simultaneously possess
localized states as in disordered media.\cite{Negro2004} The
simplest and most well-studied systems consisting of only two
structural building blocks are Fibonacci and Thue-Morse chains,
the former one being a quasicrystal. Photonic
crystals are known to allow a strong enhancement of the
light-matter interaction, particularly, if the material system has
elementary excitations and the light frequency is tuned to the
resonant frequency of these excitations (the so-called resonant
photonic crystals). In such systems the normal light waves are
polaritons. The resonant properties of elementary excitations in
quasiperiodic multilayered structures have been studied for
plasmons and spin waves, see the review
[\onlinecite{Albuquerque2003}], as well as for embedded organic
dye molecules.\cite{deych2009} Whereas resonant periodic
structures based on quantum wells (QWs) are widely investigated,
both theoretically and
experimentally,\cite{Ivchenko1994,Merle1996,koch1996mqw,Merle1997,
Haas1998,Ivchenko2000pss,Hayes1999,
Prineas2000,Deych2000,Pilozzi2004,Voronov2004} aperiodic
long-range ordered multiple QWs (MQWs) have attracted attention
quite recently.\cite{Poddubny2008prb,Hendrickson2008,Werchner2009}

In Ref.~\onlinecite{Poddubny2008prb} we have formulated the
resonant Bragg condition for the Fibonacci MQWs. We have shown
that the MQW structure tuned to this condition exhibits a
superradiant behaviour, for a small number $N$ of wells, and
photonic-crystal-like behaviour, for large values of
$N$. Moreover, in order to describe the light propagation in the
infinite Fibonacci MQWs we have applied a two-wave approximation
and derived equations for the edges of the two wide
exciton-polariton band gaps (or pseudo-gaps) where the light waves
are strongly evanescent. The fabrication and characterization of
light-emitting one-dimensional photonic quasicrystals based on
excitonic resonances have been reported in
Refs.~\onlinecite{Hendrickson2008,Werchner2009}. The measured
linear and nonlinear reflectivity spectra as a function of
detuning between the incident light and Bragg wavelength are in
good agreement with the theoretical calculations based on the
transfer-matrix approach, including the existence of a structured
dip in the pronounced superradiant spectral maximum.

In this work we further develop the theory of aperiodic MQWs with
particular references to the Fibonacci and Thue-Morse sequences. In
Secs. \ref{sec:def} and \ref{sec:superradiant} we define the
systems under study, present the results of the exact
transfer-matrix computation in the superradiant and
photonic-crystal regimes and make their general analysis. In Sec.
\ref{sec:2w} we apply the two-wave approximation to derive
analytical formulas for the light reflection and transmission
coefficients. Comparison with the exact computational results
shows that the approximate description is valid in a surprisingly
wide range of the light frequency $\omega$, the number $N$ of QWs,
and the value of nonradiative decay rate of a two-dimensional
exciton. In the close vicinity to the exciton resonance frequency
$\omega_0$ the two-wave approximation is completely invalid.
We show in Sec. \ref{sec:scaling} that in this region both the
studied aperiodic structures demonstrate scaling invariance and
self-similarity of optical properties. The main results of the
paper are briefly summarized in Sec~\ref{sec:conl}. In Appendix
the consistency of the two-wave approximation is questioned in
terms of the perturbation theory going beyond this approximation.

\section{Basic definitions}\label{sec:def}

Here we present the definitions of the aperiodic MQW chains
considered in this work. The structure consists of $N$
semiconductor QWs embedded in the dielectric matrix with the
refractive index $n_b$. Each QW is characterized by the exciton
resonance frequency $\omega_0$, exciton  radiative decay rate
$\Gamma_0$ and nonradiative  damping $\Gamma$. We neglect the
dielectric contrast assuming the background refractive index of a
QW to coincide with $n_b$. The center of the $m$-th QW ($m=1\ldots
N$) is located at the point $z=z_m$, and the points $z_m$ form an
aperiodic lattice. Three ways to define a one-dimensional
deterministic aperiodic lattice are based on the substitution
rules,\cite{lin1995++++} analytical expression for the spacings
between the lattice sites,\cite{Azbel1979} and the cut-and-project
method.\cite{Kumar1986,Janot}

We focus on the binary sequences where the interwell spacing takes
on two values, $a$ or $b$. Such structures can be associated with
a word consisting of the letters $\A$ and $\B$, where each letter
stands for the corresponding barrier. The QW arrangement is
determined by the substitutions acting on the segments $\A$ and
$\B$:
\begin{gather}\label{eq:gensubs}
\A\to \sigma(\A)=\A_1\A_2\ldots\A_{\alpha+\beta}\:,\\
\B\to \sigma(\B)=\B_1\B_2\ldots\B_{\gamma+\delta}\:. \nonumber
\end{gather}
Each of the letters $\A_k$ and $\B_k$ in the right-hand side of
Eq.~\eqref{eq:gensubs} stands for $\A$ or $\B$, $\alpha$ and
$\beta$ denote the number of letters $\A$ and $\B$ in $\sigma(A)$,
$\gamma$ and $\delta$ are the numbers of $\A$ and $\B$ in
$\sigma(\B)$, respectively.\cite{Xiujun1997} The scattering
properties of the QW sequence are described by the structure
factor
\begin{gather}\label{eq:f}
f(q) =\lim_{N \to \infty} f(q,N)\:,\\
f(q,N) = \frac{1}{N} \sum_{m = 1}^{N}\e^{2 \i q z_m}\:.\label{eq:fn}
\end{gather}
Under certain conditions\cite{Luck1993,lin1995++++} for the
numbers $\alpha$, $\beta$, $\gamma$, $\delta$ the structure
defined by Eq.~\eqref{eq:gensubs} is a quasicrystal, so that, in
the limit $N\to \infty$, the structure factor \eqref{eq:fn}
consists of $\delta$-peaks responsible for the Bragg diffraction
and characterized by two integers $h$ and $h'$,
\begin{gather}\label{eq:f2}
f(q) = \sum\limits_{h,h' = - \infty}^{\infty} \delta_{2q,G_{hh'}}
f_{hh'}\:,\\
G_{hh'}=\frac{2\pi}{\bar d}\left(h + \frac{h'}{t}\right)\
.\label{eq:Ghh}
\end{gather}
The parameter $t$ in Eq.~\eqref{eq:Ghh} is related by
\begin{equation}
 t = 1 + \frac{N_\B}{N_\A} \label{eq:t}
\end{equation}
with the numbers $N_\B, N_\A$ of the blocks $\B$ and $\A$ in the
infinitely extending lattice. The value of $t$ in \eqref{eq:t} can
be also expressed as $t = 1 + (\lambda_1 - \alpha)/\gamma$, where
$\lambda_1 =( l + \sqrt{l^2 + 4n})/2$, $l = \alpha + \delta$ and
$n = \beta \gamma - \alpha \delta$; for the quasicrystals $n$ must
be equal to $\pm 1$.\cite{kolar1993prb} The length $\bar{d}
=(a-b)/t + b$ in Eq.~\eqref{eq:Ghh} is the mean period of the
aperiodic lattice. In the periodic case where $a \equiv b \equiv
\bar{d}$, the diffraction vectors reduce to a single-index set
$G_h=2\pi h/\bar d$ with the structure-factor coefficients
$|f_h|=1$. For $a \ne b$ and irrational values of $t$, the
diffraction vectors \eqref{eq:Ghh} fill the wavevector axis in a
dense quasicontinuous way and the values of $|f_{hh'}|$ lie inside
the interval (0,1). Note that, within the uncertainty $\sim (N
\bar{d})^{-1}$, the symbol $\delta_{2q,G_{hh'}}$ in
Eq.~\eqref{eq:f2} is the Kronecker delta: $\delta_{2q,G_{hh'}}=1$
when $2q = G_{hh'}$ and zero when $2q$ is detuned from the Bragg
condition. The structure factor\cite{kolar1993} defined without
the prefactor $1/N$ in Eq. \eqref{eq:fn} is obtained by the
replacement of the Kronecker symbol in Eq.~\eqref{eq:f2} by the
functional $(2 \pi/\bar{d}) \delta(2q-G_{hh'})$.

The most famous one-dimensional quasicrystal is the Fibonacci
sequence $\A\B\A\A\B\A\B\A\ldots$ determined by the substitutions
\cite{Janot}
\begin{equation}\label{eq:Fibonacci}
\A\to \A\B,\B\to \A\:.
\end{equation}
For the canonical Fibonacci lattice the ratios $N_\A / N_\B$ and
$a/b$ are both equal to the golden mean,  $\tau = (\sqrt{5}+1)/2$.
The noncanonical Fibonacci structures with $a/b \ne \tau$ are
considered in Ref. \onlinecite{Werchner2009} and are beyond  the
scope of this paper.

 The substitution rule
\eqref{eq:Fibonacci} can be generalized in many ways to provide
other types of 1D quasicrystals. It has been proved in
Refs.~\onlinecite{lin1995++++}, \onlinecite{lin1995} that any
binary 1D quasicrystal can be obtained by substitutions composed
of different elementary inflations, e.g.,
\begin{equation}\label{eq:FibonacciGen}
\A\to \A^n\B,\:\:\B\to \A, \quad n=1,2\ldots\:.
\end{equation}

For arbitrary values of $\alpha$, $\beta$, $\gamma$ and $\delta$
the structure defined by Eq. (\ref{eq:gensubs}) does not form a
quasicrystal. For example, the substitution
\begin{equation}\label{eq:ThueMorse}
\A\to \A\B,\B\to \B\A
\end{equation}
defines the Thue-Morse lattice $\A\B\B\A\B\A\A\B\ldots$ with a
singular continuous structure factor and the mean period $\bar
d=(a+b)/2$.\cite{Savit1988} For the Thue-Morse QW structure the
function $f(q,N)$ in Eq. \eqref{eq:fn} tends to zero when $N \to
\infty$ as a power of $N$ at any $q$ except  certain singular
values. The latter form a series
\begin{equation}\label{eq:Gh}
2q = G_h = \frac{\pi h}{ \bar{d} },\quad h=0,\pm 1\ldots
\end{equation}
with the structure factor given by\cite{Liviotty1996}
\begin{equation}
|f_h^{({\rm TM})}| = \cos^2\left(\frac{\pi a h}{2 \bar{d}}\right) =
\cos^2\left(\frac{\pi b h}{2 \bar{d}}\right)\:.
\end{equation}

The resonant Bragg condition\cite{Poddubny2008prb} for both the
Fibonacci and Thue-Morse structures is formulated as
\begin{equation}\label{eq:resBragg}
\frac{\omega_0 n_b}{c} = \frac{G}{2}\:,
\end{equation}
where $G$ stands for $G_{hh'}$ in the Fibonacci case and for
$G_{h}$ in the Thue-Morse case, see Eqs.~\eqref{eq:Ghh} and
\eqref{eq:Gh}, respectively. Of course, one can impose a similar
condition for non-singular wavevectors contributing to the
structure factor of the Thue-Morse sequence. Since in this case
the value $f(q,N)$ decreases with increasing $N$ the corresponding
system is far from being an efficient exciton-polaritonic
structure. This is the reason why we do not consider here, e.g.,
the period-doubling sequence
$\A\B\A\A\A\B\A\B\ldots$\cite{Liviotty1996} determined by the rule
$\A\to \A\B,\B\to \A\A$, which has no Bragg peaks except for the
trivial one at $q=0$.

For the sake of completeness, we also analyze a slightly
disordered structure with the long-range order maintained and the
QW positions defined by
\begin{equation}\label{zmdis}
z_m = m \bar d + \delta z_m\:,
\end{equation}
where the deviation $\delta z_m$ is randomly distributed and
defined by the vanishing average, $\langle \delta z_m \rangle =0$,
and the dispersion $\sigma_z^2 = \langle \delta z_m^2\rangle$. The
structure factor $f(q) = \lim_{N \to \infty} f(q,N)$ of such a
lattice averaged over the disorder realizations has the form
\begin{equation}\label{eq:DW}
\langle f(q) \rangle = \sum\limits_{h}\delta_{2q,G_h}\e^{-(q\sigma_z)^2/2}
,\quad G_h=2\pi h/\bar d\:.
\end{equation}
The dispersion of $f(q,N)$ tends to zero with $N \to \infty$, and
Eq.~\eqref{eq:DW} provides a good estimation of the structure
factor for any fixed disorder realization whenever $N \gtrsim 10$.
The long-ranged correlations of QW positions are preserved by
\eqref{zmdis}, and the Bragg diffraction is possible with the same
diffraction vectors as in the periodic lattice. However, the
structure-factor coefficients drop drastically with the growth of
$\sigma_z$. The exponential factor in \eqref{eq:DW} is equivalent
to the Debye-Waller factor caused by the thermal motion of atoms
in a crystalline lattice.\cite{KittelIntro}

Since the geometry of MQW structures under study is now described
and the resonant Bragg condition is imposed we proceed to the
optical reflection spectra.

\begin{figure}[h!]
\includegraphics[width=0.49\textwidth]{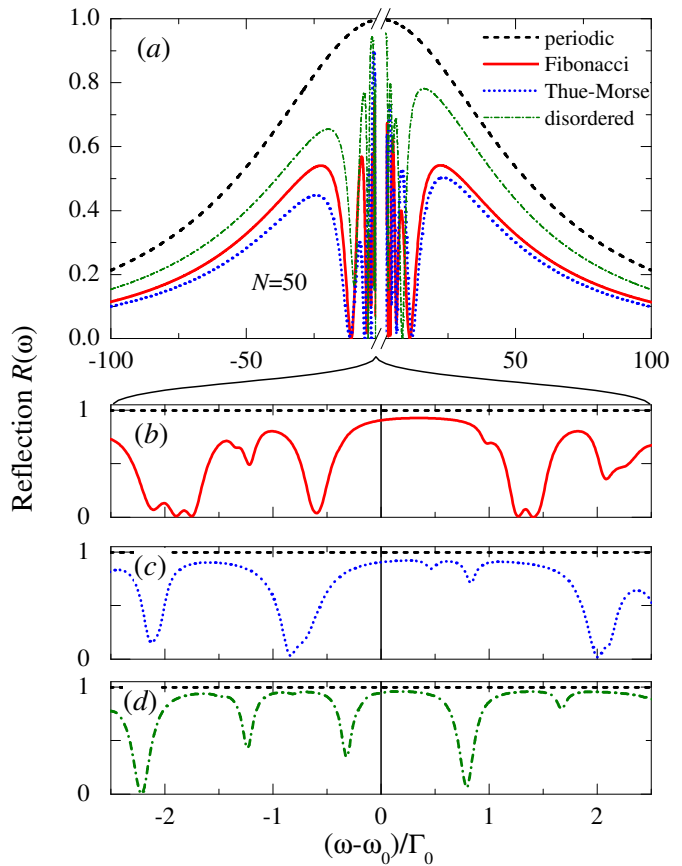}
 \caption{(Color online)
Reflection spectra calculated for four QW structures, each containing $N = 50$ wells and tuned to the resonant Bragg condition $2 \bar{d} = \lambda(\omega_0)$:
periodic structure, with $a \equiv b \equiv \bar{d}$ (dashed);
Fibonacci chain, with $a / b = \tau$ (solid curve);  Thue-Morse sequence with
$a/b = 3/2$ (dotted); and weakly disordered periodic
with $\sigma_z = \lambda_0/20$ (dashed-and-dotted).
Note the break on abscissa axis on the panel (a) around $\omega=\omega_0$.
Panels (b), (c) and (d) show the same spectra in larger scale of the variable $(\omega -
\omega_0)/\Gamma_0$.
Calculated for
$\hbar \Gamma_0 = 50~\mu$eV, $\hbar\omega_0 = 1.533$~eV, $\Gamma = 0.1
\Gamma_0$.
}\label{fig:1}
\end{figure}

\section{Two regimes in optical reflection from resonant Bragg structures}\label{sec:superradiant}
Two different regimes have been revealed in optical reflection from
the periodic resonant Bragg QW structures.\cite{Ivchenko2000pss,Cho2002} For
small enough numbers of QWs, $N \ll \sqrt{\omega_0/\Gamma_0}$
(superradiant regime), the optical reflectivity is described by a
Lorentzian with the maximum value $[N \Gamma_0/ (N \Gamma_0 +
\Gamma)]^2$ and the halfwidth $N \Gamma_0 + \Gamma$. For a large
number of wells, $N > \sqrt{\omega_0/\Gamma_0}$ (photonic crystal
regime), the reflection coefficient is close to unity within the
exciton-polariton forbidden gap and exhibits an oscillatory
behaviour outside the gap.
The calculations presented below
demonstrate the existence of similar two regimes for the
deterministic aperiodic structures.
\subsection{Superradiant regime}

The numerical calculation of reflection spectra is carried out
using the standard transfer matrix technique\cite{Ivchenko2005}
given with more details in Sec.~\ref{sec:scaling}.
Figure~\ref{fig:1} presents the reflectivity $R_N(\omega)$
calculated for the light normally incident from the left
half-space $z<0$ upon four different 50-well structures. All the
four, namely, the Fibonacci, Thue-Morse, periodic and distorted
periodic structures, are tuned to satisfy the Bragg resonant
condition (\ref{eq:resBragg}) which can be rewritten as
$\lambda(\omega_0) = 2 \bar{d}$, where $\lambda(\omega) = 2 \pi
c/(\omega n_b)$. This means that, for the Fibonacci structure, the
value $G$ in Eq.~(\ref{eq:resBragg}) is set to $G_{hh'}$ with
$h=1,h'=0$ and, therefore, for the four structures $G = 2
\pi/\bar{d}$ and their optical properties can be conveniently
compared.

One can see from Fig.~\ref{fig:1} that the condition
(\ref{eq:resBragg}) leads to high reflectivity of not only the
periodic and quasicrystalline Fibonacci
chains\cite{Poddubny2008prb} but also the Thue-Morse and
slightly-disordered periodic structures. In the region $|\omega -
\omega_0| > 20 \Gamma_0$, far enough from the exciton resonance
frequency, the four spectra have similar Lorentzian wings with the
halfwidth of the order of $N \Gamma_0$ indicating the existence of
a superradiant exciton-polariton mode. The magnitude of the wings
is governed by a modulus of the structure-factor coefficient,
$|f_G|$. For the chosen structures this value runs from $|f_G| =
1$ (periodic structure) and $|f_G| = 0.95$ (distorted periodic) to
$|f_G| = 0.70$ (Fibonacci) and $|f_G| = 0.65$ (Thue-Morse). The
spectral wings in Fig.~\ref{fig:1} decline monotonously with
decreasing $|f_G|$. In addition it should be mentioned that, for
the Fibonacci QW structure tuned to $G_{hh'}$ with $h=1, h'=1$ and
analyzed in Ref.~\onlinecite{Poddubny2008prb}, the
structure-factor coefficient is $|f_G| \approx 0.9$ and the spectral
wings in reflectivity are raised as compared with those for the
Fibonacci structure tuned to $G_{1,0}$.

In the frequency region around $\omega_0$ the reflection spectra
from the non-periodic structures show wide dips where the
reflection coefficient oscillates with the period of oscillations
decreasing as $\omega$ approaches $\omega_0$. As shown below, see
also Ref.~\cite{Poddubny2008prb}, the spectral dip naturally appears
for a multi-layered deterministic system tuned to a Bragg
diffraction vector with the structure-factor coefficient $f_G$
smaller than unity, and it widens as the value of $1 - |f_G|$
increases.

For small values of the exciton nonradiative damping rate $\Gamma$
(lying beyond experimentally available values), the fine structure
of optical spectra in the narrow resonance region ranged over few
$\Gamma_0$ is very intricate, see
Figs.~\ref{fig:1}(b)--\ref{fig:1}(d). All the considered aperiodic
structures possess a narrow middle stop-band embracing the exciton
resonance $\omega_0$. In particular, for the Fibonacci QW
structure this stop-band is located between $\omega_0 - 0.4
\Gamma_0$ and $\omega_0 + 0.9 \Gamma_0$. The  spectral properties
in the frequency range $|\omega-\omega_0| \sim \Gamma_0$ for
$\Gamma \ll \Gamma_0$ are discussed in Sec.~\ref{sec:scaling} in
more details.

\begin{figure}[h!]
\includegraphics[width=0.49\textwidth]{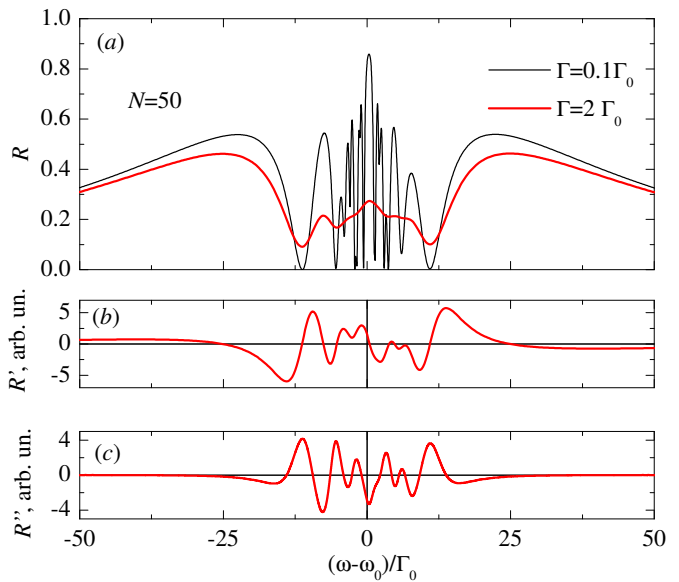}
 \caption{(Color online)
Differential reflection spectra calculated for Fibonacci QW
structures containing $N = 50$ wells. Panel (a) shows reflection spectrum
$R(\omega)$ calculated for $\Gamma = 0.1 \Gamma_0$ (thin curve) and
$\Gamma =2 \Gamma_0$ (thick curve). Panels (b) and (c) demonstrate
the first- and second-order differential spectra $R'(\omega)$ and
$R''(\omega)$ in arbitrary units for $\Gamma = 2\Gamma_0$. Other
parameters are the same as in Fig.~\ref{fig:1}.} \label{fig:1p}
\end{figure}

\begin{figure}[h!]
\includegraphics[width=0.49\textwidth]{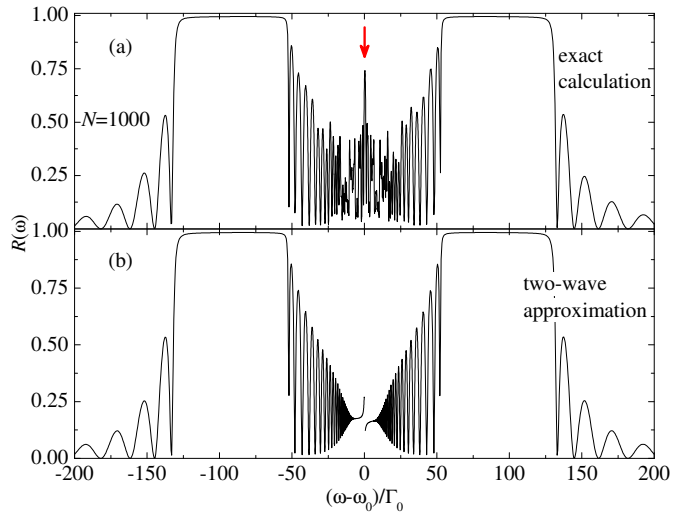}
 \caption{(Color online) Reflection spectra calculated for a
Fibonacci QW structure containing 1000 wells. The parameters used
are the same as those in Fig.~\ref{fig:1} except for the nonradiative decay
which now is $\Gamma = 0.2 \Gamma_0$.  Upper and lower panels correspond,
respectively, to the exact calculation and calculation in the two-wave
approximation. Vertical arrow at $\omega = \omega_0$ in panel (a) indicates
the narrow reflectivity stop-band, which cannot be described by the two-wave
approximation.} \label{fig:2}
\end{figure}

In realistic semiconductor QWs the nonradiative decay rate is
larger than (or comparable with) $\Gamma_0$, and the majority of
spectral fine-structure features are
smoothed.\cite{Hendrickson2008} The influence of the nonradiative
damping on the reflectivity from the Fibonacci MQWs is illustrated
in Fig.~\ref{fig:1p}. Thin curve on the upper panel is the same as
that on Fig.~\ref{fig:1} and calculated for $\Gamma = 0.1\Gamma_0$
while the thick curve corresponds to more realistic value $\Gamma
= 2 \Gamma_0$. One can trace the smoothing of sharp spectral
features with increasing $\Gamma$. However, some of these features
may still be resolved by means of the differential spectroscopy
widely used in studies of bulk crystals and low-dimensional
structures.\cite{Makhniy2006} Panels (b) and (c) of
Fig.~\ref{fig:2} present the first and second derivatives
$R'(\omega) \equiv \partial R(\omega)/ \partial \omega$ and
$R''(\omega)$, respectively. The differential spectra allow one to
enhance the spectral peculiarities poorly resolved in the ordinary
spectrum of Fig.~\ref{fig:1p}(a).


\subsection{Photonic-quasicrystal regime}

\begin{figure}[h!]
\includegraphics[width=0.49\textwidth]{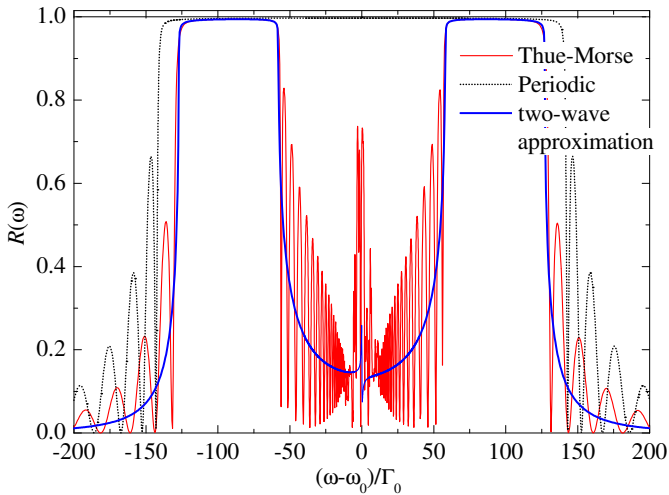}
 \caption{(Color online)
Reflection spectra calculated for  Thue-Morse and periodic QW structures.
Thin solid curve corresponds to the exact calculation for
$N=1000$ Thue-Morse QWs, thick solid curve is calculated in the two-wave
approximation for
$N \to \infty$. Dotted curve represents the $N=1000$ periodic MQWs with the
period $d = \lambda(\omega_0)/2$.
The parameters used are the same as those in Fig.~\ref{fig:2}.} \label{fig:3}
\end{figure}

The superradiant regime holds up to $N \sim 100$ and is followed
by a saturation with the further increase in the number of
QWs.\cite{Cho2002} In this subsection we study very long QW chains
with $N \gg \sqrt{\omega_0/\Gamma_0}$. The calculations for
$N=1000$ are presented in Fig.~\ref{fig:2} for Fibonacci MQWs and
Fig.~\ref{fig:3} for Thue-Morse MQWs. Figure~\ref{fig:2} allows a
clear interpretation of the spectral properties of excitonic
polaritons. Two symmetric stop-bands\cite{Poddubny2008prb}
standing out between numerous sharp maxima and minima are clearly
seen in the spectra. Figure~\ref{fig:2}(b) shows the spectrum
calculated in the two-wave approximation taking into account only
three terms in the sum (\ref{eq:f2}), namely, the terms with $2q =
\pm G_{1,0}$ and $2q = 0$, see Sec.~\ref{sec:2w} for details.
Comparing Figs.~\ref{fig:2}(a) and \ref{fig:2}(b) we conclude that a
lot of spectral features are reproduced as the interference
fringes in the approximated spectrum. However, this approximation
lacks an adequate description of the reflection spectrum around
the exciton resonance frequency. The middle stop-band at $\omega =
\omega_0$ found in Fig.~\ref{fig:1}(b) reveals itself also in
Fig.~\ref{fig:2}(a) where it is indicated by a vertical arrow.

Thin solid curve in Fig.~\ref{fig:3} illustrates the exact
reflectivity calculation for the Thue-Morse structure with
$N=1000$. Thick solid curve is calculated in the two-wave
approximation in the limit $N \to \infty$ so that all the
interference fringes are smoothed due to the finite value of
$\Gamma \to + 0$. We have checked for the Thue-Morse sequence
that, for $N=1000$, the two-wave approximation works well outside
the narrow interval $|\omega - \omega_0| \sim \Gamma_0$. The fine
features around $\omega_0$ are again beyond the scope of the
two-wave approximation. The spectrum for 1000 periodic QWs, dotted
curve in Fig.~\ref{fig:3}, is presented for comparison in order to
emphasize similarities and differences between the optical spectra
of periodic and aperiodic systems under consideration.

Figures \ref{fig:1}-\ref{fig:3} form a computational data base for
the physical interpretation of the reflection spectral shapes.
This can be done in terms of the two-wave approximation
(Sec.~\ref{sec:2w}) and scaling and self-similarity
(Sec.~\ref{sec:scaling}).

\section{Optical spectra in the two-wave approximation}\label{sec:2w}
In this section we apply the two-wave approximation for derivation
of the exciton-polariton dispersion and the reflectivity spectra
of the aperiodic resonant Bragg MQW structures. The derivation is
performed for the particular case of a canonic Fibonacci chain but
the results can be straightforwardly generalized to
noncanonical quasicrystal sequences and other
deterministic aperiodic multi-layered structures.

The electric field of the light wave propagating in the MQW
structure satisfies the following wave equation\cite{Ivchenko2005}
\begin{equation}\label{eq:maxwell}
\left(-\frac{d^2}{dz^2}-q^2\right) E(z) = \frac{2 q \Gamma_0}{\omega_0 -
\omega - {\rm i} \Gamma}\sum_{m} \delta(z - z_m)E(z_m)\:.
\end{equation}
Assuming the resonant Fibonacci MQWs to contain a sufficiently
large number $N$ of wells we replace the structure factor $f(q,N)$
by its limit $f(q)$ given by Eq.~(\ref{eq:f2}). This allows one to
approximate solutions of Eq.~\eqref{eq:maxwell} as a superposition
of the ``Bloch-like" waves
\begin{equation}\label{eq:bloch}
E_K(z) = \sum\limits_{h,h'} \e^{\i (K - G_{hh'})z} E_{K-G_{hh'}}\:.
\end{equation}
As distinct from the true Bloch functions in a periodic system,
here the countable set of vectors $G_{hh'}$ is enumerated not by
one but by two integer numbers, $h$ and $h'$, see
Eq.~(\ref{eq:Ghh}). Substituting $E_K(z)$ into
Eq.~\eqref{eq:maxwell}, multiplying each term by $\exp[- \i (K -
G_{hh'})z]$ with particular $h,h'$ and integrating over $z$ we
obtain
\begin{equation}\label{eq:fourier}
\begin{split}
[q^2 - (K - &G_{hh'})^2 + 2q \xi] E_{K - G_{hh'}} +\\&2q \xi \sum\limits_{gg'}
f^*_{g -h, g' - h'} E_{K - G_{gg'}} = 0\:, 
\end{split}
\end{equation}
where
\begin{equation}\label{eq:chi}
 \xi(\omega)=\frac{\Gamma_0}{{\bar d}(\omega_0 - \omega- {\rm i} \Gamma)}\:.
\end{equation}
Note that, throughout this paper, we focus on a frequency
region $|\omega - \omega_0| \ll \omega_0$ around the exciton
resonance.

In accordance with Eq.~(\ref{eq:resBragg}) we consider the
Fibonacci QW structure tuned to the Bragg resonance
\begin{equation}
\frac{\omega_0}{c} n_b \equiv q_0 = \frac{G_{h,h'}}{2} \:.
\end{equation}
In the two-wave approximation, only two space harmonics $K$ and
$K' = K - G_{hh'} = K - 2 q_0$ are taken into account in the
superposition (\ref{eq:bloch}). A necessary but not sufficient
condition for validity of this approximation is the inequality
\begin{equation} \label{crit}
\left\vert q_0 - K \right\vert \ll q_0\ .
\end{equation}
Then the infinite set (\ref{eq:fourier}) is reduced to a system of
two coupled equations
\begin{eqnarray}\label{eq:w2}
&&(q - K + \xi) E_K  +  \xi f^{*}_{hh'} E_{K'} = 0\:, \\
&&\xi f_{hh'} E_{K} + (q + K - 2q_0 + \xi) E_{K'} = 0\:. \nonumber
\end{eqnarray}
The two eigenvalues $K$ corresponding to the frequency $\omega$
are given by
\begin{equation} \label{KQ}
K^{(\pm)} = q_0 \pm Q\:,\: Q = \sqrt{(\xi + q - q_0)^2 - \xi^2 |f_{hh'}|^2}\:.
\end{equation}
The criterion (\ref{crit}) is then rewritten in the form
\begin{equation} \label{crit2}
{\rm max} \{ |\omega_0 - \omega|, \Gamma\} \gg \Gamma_0
\frac{ \sqrt{1 - |f_{hh'}|^2} }{ h + h'/\tau }\ .
\end{equation}

At $f_{hh'} \to 1$ this dispersion equation reduces to that for
the periodic resonant Bragg MQWs\cite{Ivchenko2000pss}
\[
Q = q_0 \sqrt{\left( \frac{\omega - \omega_0}{\omega_0} \right)^2 -
\Delta^2 \frac{\omega - \omega_0}{\omega - \omega_0 + {\rm i} \Gamma}}\:,
\]
where
$$
\Delta = \sqrt{\frac{2 \Gamma_0 \omega_0}{\pi}}\ .
$$

The edges, $\omega_{\rm out}^{\pm}$ and $\omega_{\rm in}^{\pm}$,
of two symmetrical band gaps in the Fibonacci QW structure are
obtained from Eq.~(\ref{KQ}) by setting $K^{(\pm)} = q_0$ or,
equivalently, $Q = 0$. The result reads\cite{Poddubny2008prb}
\begin{gather}\label{eq:ws}
\omega_{\rm out}^{\pm} = \omega_0 \pm \Delta \sqrt{\frac{1 + |f_{hh'}|}{ 2\:(h+h'/\tau)}}\:,\\
\omega_{\rm in}^{\pm} = \omega_0 \pm \Delta \sqrt{\frac{1 -
|f_{hh'}|}{ 2\:(h + h'/\tau)}} \nonumber \:.
\end{gather}
As shown in Appendix, for the exciton-polariton waves at the
bandgap edges located at the point $K = q_0$, an admixture of
other space harmonics has no remarkable influence and these edges
are well-defined for the resonant QW Fibonacci chains. For each
frequency lying inside the interval between the edges $\omega_{\rm
in}^{-}$ and $\omega_{\rm in}^{+}$, the two-wave approximation
gives two linearly independent solutions $K^{(\pm)} = q_0 \pm Q$
with nonzero $Q$. An exciton-polariton wave induced by the initial
incoming light wave is a superposition of these two Bloch-like
solutions. They can be coupled by the diffraction wavevector
$G_{gg'}$ satisfying the condition $2 Q = G_{gg'}$. If the
corresponding structure-factor coefficient $f_{gg'}$ is remarkable
one should include into consideration mixing of the waves
$K^{(\pm)}$ which complicates the comparatively simple description
of exciton polaritons. In the approximate approach we will ignore
the diffraction-induced mixing between the waves $K^{(\pm)}$ and
analyze the validity of this description by comparing the exact
and two-wave calculations.

In order to derive an analytical expression for the reflection
coefficient from an $N$-well chain sandwiched between
the semiinfinite barriers (material B) we write the field in the
three regions, the left barrier, the MQWs and the right barrier,
as follows
\begin{equation} \label{eq:field}
E(z) = \begin{cases}
e^{{\rm i}qz} + r_N e^{-{\rm i}qz} \quad &(z < 0)\:,\\
E_+ {\rm e}^{{\rm i} Q z} ({\rm e}^{{\rm i} q_{\mbox{}_{0}} z} + \zeta_+ {\rm e}^{-{\rm i} q_{\mbox{}_{0}} z})
+ \\\quad E_- {\rm e}^{-{\rm i} Q z} ({\rm e}^{{\rm i} q_{\mbox{}_{0}} z} + \zeta_- {\rm e}^{-{\rm i} q_{\mbox{}_{0}} z})
\quad &(0 < z < N \bar d)\:, \\
t_N e^{i q (z - N \bar{d})} \quad &(N \bar d < z)\:.
\end{cases}
\end{equation}
Here $r_N$ and $t_N$ are the reflection and transmission amplitude
coefficients, $E_{\pm}$ are the amplitudes of the ``Bloch-like"
solutions, and
\[
\zeta_{\pm} = - \frac{\xi f_{hh'}}{q - q_0 \pm Q + \xi}\:.
\]
Values of $r_N, t_N, E_+, E_-$ are related by imposing the
boundary conditions which are continuity of the electric field
$E(z)$ and its first derivative $dE(z)/dz$ at the points $z_1 = 0$
and $z_N$. If the number of wells $N$ coincides with $N = F_j +
1$, where $F_j$ is one of the Fibonacci numbers, then the product
$G_{hh'} z_N$ differs from an integer multiply of $2 \pi$ by a
negligibly small value, $\delta_N= - G_{hh'}s (1 - \tau)^{j-1}
\tau^{-2}$. In this case the phase factor $\exp{({\rm i}
G_{hh'}z_N)}$ can be replaced by unity and the straightforward
derivation results in surprisingly simple expressions for the
reflection coefficient,
\begin{equation}\label{eq:rSR}
r_N = \frac{\xi f_{hh'}} {q_0 - q - \xi - {\rm i} Q \cot{(Q N \bar{d})} }
\:,
\end{equation}
and for the transmission coefficient, ${t_N=-{\rm i} r_N Q/[\xi
f_{hh'}\sin{(Q N \bar{d})}]}$.

Numerical calculation demonstrates that in the region
\eqref{crit2} the two-wave approximation is valid even for the
Fibonacci structures with  $N \ne F_j +1$ provided that $N \gtrsim
20$ and the mesoscopic effects are reduced. Moreover,
Eq.~\eqref{eq:rSR} for the reflection coefficient can be applied
to other deterministic aperiodic systems including the Thue-Morse
and weakly disordered periodic structures. It suffices for the
structure to be characterized by a single value of the structure
factor at particular vector which can play the role of the Bragg
diffraction vector. In the crystalline case where $f_{hh'}\equiv
1$ analogous approximation describes the nuclear resonant
scattering of $\gamma$-rays.\cite{kagan1999}

The two-wave approximation can be generalized to allow for the
dielectric contrast, i.e., the difference between the dielectric
constant of the barrier, $n_b^2$, and the background dielectric
constant of the QW, $n_a^2$. In a structure with $n_a \ne n_b$ the
stop-band exists even neglecting the exciton effect, $\Gamma_0 =
0$. The excitonic resonance leads to the splitting of this single
stop-band to two ones. In the periodic case the highest
reflectivity is reached when the two stop-bands touch each other
and form a single exciton-polariton gap.  This is the effective
Bragg condition for the periodic structure with the dielectric
contrast.\cite{Voronov2004} In the Fibonacci case when
$|f_{hh'}|<1$ the stop bands never touch each other and the Bragg
condition means that the sum of  their widths reaches a maximum.
For the realistic case of a small contrast, $|n_a - n_b|\ll n_a,
n_b$, this condition is equivalent to the tuning of the exciton
resonance frequency $\omega_0$ to one of the edges of the
stop-band found at $\Gamma_0=0$, similarly to the condition for
the periodic structures.\cite{Voronov2004,Ivchenko2006en}. Note
that the reflectivity spectrum taken from the Bragg MQW structure
with the dielectric contrast is always asymmetric.

For periodic resonant Bragg MQWs, in the superradiant regime $N
\ll \sqrt{\omega_0/\Gamma_0}$, Eq.~(\ref{eq:rSR}) is readily
transformed to the well-known result\cite{Ivchenko1994}
\begin{equation}\label{eq:rSR2}
 r_N(\omega)=\frac{{\rm i} N \Gamma_0}{\omega_0 - \omega - \i (N\Gamma_0+\Gamma)}\:.
\end{equation}
The pole at $ \omega = \omega_0 - \i (N\Gamma_0 + \Gamma)$ is the
eigenfrequency of the superradiant mode. In general, the
eigenfrequencies $\omega^{(l)}$ of a MQW structure are represented
by zeros of the denominator in Eq.~(\ref{eq:rSR}). Since the
structure is open the eigenfrequencies are complex even in the
absence of nonradiative damping, $\Gamma = 0$. Values of
$\omega^{(l)}$ lying in the region
\begin{equation} \label{crit3}
|\omega - \omega_0| \ll \frac{\Delta\sqrt{1-|f_{hh'}|}}{\sqrt{h + h'/\tau} }
\end{equation}
but outside the narrow interval (\ref{crit2}) can be easily found
taking into account that, in this region, the difference $q_0 - q$
in Eqs.~(\ref{KQ}), (\ref{eq:rSR}) can be neglected as compared
with $\xi$ so that one has $Q = \xi \sqrt{1-|f_{hh'}|^2}$ and
\begin{gather}
\omega^{(l)}=\omega_0 - \i \Gamma - \i N \Gamma_0 \frac{\sqrt{1 -
|f_{hh'}|^2}}{\arctanh (\sqrt{1-|f_{hh'}|^2})+ {\rm i} \pi l}\
,\nonumber\\ l=0, \pm 1, \pm 2\ldots \label{eq:allfreqs}
\end{gather}
Equation~(\ref{eq:allfreqs}) determines at $l = 0$ the frequency
of the superradiant mode. In the particular case $1 - |f_{hh'}|^2
\ll 1$, this frequency is given by
\begin{equation}\label{eq:SR}
\omega^{(0)} = \omega_0 - {\rm i} N \Gamma_0 \left(\frac{2}{3} +
\frac{ |f_{hh'}|^2}{3}\right)\:.
\end{equation}
\begin{figure}[h!]
\includegraphics[width=0.49\textwidth]{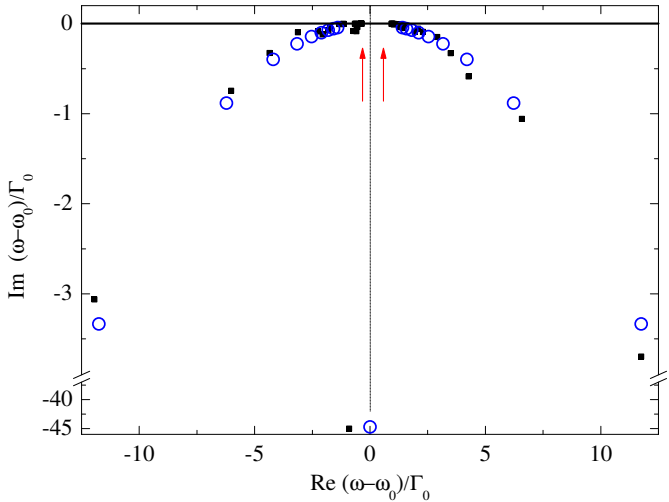}
 \caption{(Color online) Complex eigenfrequencies of exciton
polaritons in Fibonacci 56-QW structure calculated exactly (filled symbols) and in
the two-wave approximation (empty symbols). Note the break at the ordinate axis.
Vertical arrows indicate the edges of inner stop-band. Calculated for $\Gamma=0$
and other parameters same as in Fig.~\ref{fig:1}} \label{fig:4}
\end{figure}

Figure~\ref{fig:4} shows the eigenfrequencies of a $56$-QW
Fibonacci structure calculated exactly (filled symbols) and from
Eq.~\eqref{eq:allfreqs} (closed symbols). The exact calculation is
performed using the system of coupled equation for excitonic
polarization $P_m$ in the quantum well\cite{Ivchenko2005}
\begin{equation}\label{eq:oscillators}
 (\omega_0-\omega-\i\Gamma)P_{m}+\sum\limits_{m'=1}^{N}e^{\i q|z_m-z_m'|}P_{m'},\quad m=1\ldots N\:.
\end{equation}
One can see from Fig.~\ref{fig:4} that the two-wave approximation
excellently describes the superradiant mode as well as some of the
subradiant modes lying far from $\omega_0$ on the complex plane.
The approximation breaks  in the region close to $\omega_0$ where
 more sophisticated analysis is required, as presented in the
next Section.

\section{Optical spectra near the exciton resonance frequency}\label{sec:scaling}
\subsection{Trace map technique for Fibonacci and Thue-Morse quantum well
structures}
In this section we concentrate on the narrow frequency region
$|\omega - \omega_0| \sim \Gamma_0$. A convenient and powerful
tool for such study is the transfer matrix method. The
substitution rules \eqref{eq:gensubs} lead to recursive relations
for the transfer matrices providing all the essential information
about the spectral properties of MQWs.

In the following we use the notation $\F_j$ for the Fibonacci
chain containing $N=F_j$ QWs starting from the trivial chain
$\F_1$ that consists of one segment $\A$. Similarly, $\TM_j$ is
the Thue-Morse sequence with $N=2^j$ QWs starting from $\TM_0=\A$.
The transfer matrix $M_j$ through the whole structure $\F_j$ or
$\TM_j$ is given by a product of the matrices $M_{\rm QW}$,
$M_{\rm A}$, $M_{\rm B}$ of transfer through a QW and a barrier of
length $a$ or $b$, respectively, with the order established by the
chain definition \eqref{eq:gensubs}. In the basis of electric
field $E(z)$ and its derivative $- q^{-1} dE(z)/dz$ the transfer
matrices are as follows\cite{Ivchenko2005}
\begin{equation}\label{eq:Mqw}
M_{\rm QW}=\begin{pmatrix}
1&0\\ 2S&1
            \end{pmatrix}\:,\:
            S=\frac{\Gamma_0}{\omega_0-\omega-i\Gamma}\:,
\end{equation}
and  \cite{Brehm1991}
\begin{equation}\label{eq:MD}
             M_{\mathcal D}=\begin{pmatrix}
\cos{qd}&-\sin{qd}\\ \sin{qd}&\cos{qd}
            \end{pmatrix},\quad \mathcal D= {\rm A},{\rm B}; d = a, b \:.
\end{equation}
We will here restrict ourselves to the limit of zero nonradiative decay,
$\Gamma = 0$, in which case the transfer matrices are real. The
transmission and reflection spectra, $T_j(\omega)$ and
$R_j(\omega)$, are given by\cite{Brehm1991}
\begin{equation}\label{eq:TR}
 T_{j}(\omega) = 1 - R_{j}(\omega) = \frac{1}{x_{j}^2(\omega) + y_{j}^2(\omega)}\:.
\end{equation}
Here the quantities $x_j$ and $y_j$ stand for the half-trace
$(M_{j,11}+M_{j,22})/2$ and half-antitrace
$(M_{j,21}-M_{j,12})/2$ of the matrix $M_j$, respectively.

In order to reveal the behavior of exciton polaritons in aperiodic
MQWs it is instructive to calculate the polariton dispersion in
the approximants\cite{Azbel1979} of the aperiodic chains
containing the periodically repeating sequences $\F_j$ or $\TM_j$.
In such periodic systems the polariton band structure consists of
allowed minibands and forbidden gaps. The gaps are found from the
condition\cite{Ivchenko2005}
\begin{equation}\label{eq:edges}
|x_j(\omega)| > 1\:.
\end{equation}
To proceed to the analysis of the pattern of allowed and forbidden
bands we note that the half-traces $x_j$ of the substitution
sequences satisfy closed recurrence relations, also termed as
trace maps.\cite{kolar1990} For the Fibonacci and Thue-Morse
chains, the trace maps read
\begin{subequations} \label{eq:req}
\begin{align}
x_{j+1}&=2x_{j}x_{j-1}-x_{j-2}&&\text{(Fibonacci)}\:, \label{eq:req1}\\
x_{j+1}&=4x_{j-1}^2(x_{j}-1)+1&&\text{(Thue-Morse)}\:.\label{eq:req1b}
\end{align}
\end{subequations}
Consequently, the polariton energy spectrum is determined by the
general properties of nonlinear transformations \eqref{eq:req} and
the initial conditions specific for the QW transfer
matrices~\eqref{eq:Mqw}. The trace maps are  effective for
 studies of the spectral properties of deterministic aperiodic
structures.\cite{Kokmoto1986, lin1995++++}

\subsection{Scaling of band structure and transmission spectra in Fibonacci
structures}
\begin{figure}[h!]
\includegraphics[width=0.49\textwidth]{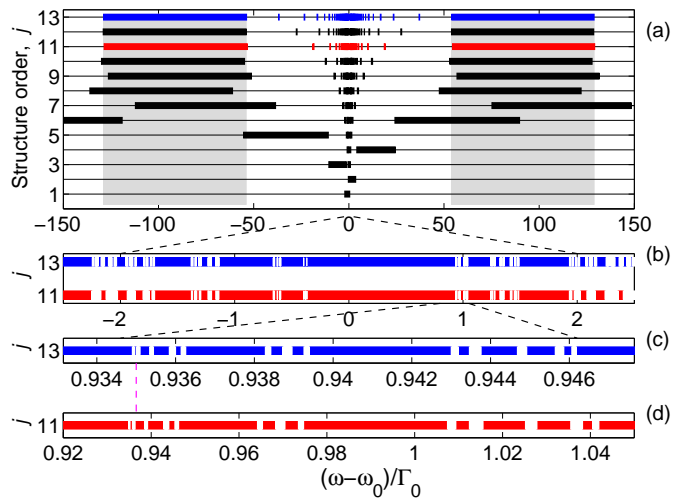}
 \caption{(Color online) (a). Exciton-polariton allowed (thin lines)
and forbidden (thick stripes) bands in periodically repeated
Fibonacci sequences of the order
$j=1 \ldots 13$. (b) Bands for $j=11$ ad $j=13$ in the spectral range  around the frequency $\omega=\omega_0$. (c),(d) Bands for $j=11$ and $j=13$, respectively, in a large scales near the frequency  $\omega=\omega_0+0.935\Gamma_0$ indicated by the vertical line.   Calculated for $\Gamma=0$
and other parameters same as in Fig.~\ref{fig:1}
}\label{fig:5}
\end{figure}

We will now analyze the band structure and the transmission spectra in
Fibonacci QW structures.
For the Fibonacci lattices the trace map \eqref{eq:req1}
possesses
an invariant\cite{Kohmoto1984physlett}
$$I=x_{j}^2+x_{j+1}^2+x_{j+2}^2-2x_jx_{j+1}x_{j+2}-1\:.
$$
In the QW
structure this invariant can be reduced to
\begin{equation} \label{eq:I}
 I(\omega)=S^2(\omega)\sin^2[qb(\tau-1)]\:.
\end{equation}
The resonant behavior of this invariant as a function of frequency
indicates that the band structure for Fibonacci QW chains may be
very complex in the region $|\omega-\omega_0|\sim \Gamma_0$.  The
band calculations are presented in Fig.~\ref{fig:5}(a), where the
black stripes and horizontal lines show, respectively, the
forbidden and allowed bands  for different values of the structure
order $j$ ranging from $j=1$ to $j =13$.
Figures~\ref{fig:5}(b)-\ref{fig:5}(d) represent this band sequence
for $j=11$ and $j=13$ in different frequency scales. Panel (a)
demonstrates that two broad band gaps are already present for 21
QWs ($j=8$). With increasing $j$ their edges very quickly converge
to the analytical values \eqref{eq:ws} shown by the gray
rectangles in Fig.~\ref{fig:5}(a). A narrow permanent middle band
gap at $-0.4\Gamma_0 \lesssim \omega - \omega_0 \lesssim
0.9\Gamma_0$ is well resolved in the scale of Fig.~\ref{fig:5}(b).

The other forbidden bands depicted in Fig.~\ref{fig:5} can be interpreted in
terms of two  formation mechanisms. The first
mechanism is related to the two-wave approximation. In this
approximation the half-trace of the transfer matrix $x_j$ reaches
minimum ($-1$) or maximum (+1) values at particular frequencies
$\omega_s$ where the reflectivity $r_N^{({\rm 2w})}$ vanishes. One
can check from Eq.~(\ref{eq:rSR}) that at these frequencies the
product $Q N \bar{d}$ is an integer number of $\pi$. In the
vicinity of $\omega_s$ one can approximate $x^{({\rm
2w})}(\omega)$ by $n_s [1 - u_s (\omega - \omega_s)^2]$, where
$n_s = \pm 1$, and $u_s$ is a positive coefficient. Near
$\omega_s$ the exact function $x_j(\omega) \equiv {\rm Tr}(M_j)/2$
differs from $x^{({\rm 2w})}(\omega)$ by the correction $\delta
x(\omega)$ which can be approximated by $n_s [c_s + v_s (\omega -
\omega_s)]$ where $c_s, v_s$ are additional constants. As a result
the behavior of the half-trace can be presented in the form
\[
x(\omega) = n_s\left[ 1 + c_s + \frac{v_s^2}{4u_s} -
u_s \left(\omega - \omega_s - \frac{v_s}{2u_s} \right)^2 \right]\:.
\]
If $c_s + (v_s^2/4u_s)$ is positive then the periodic system has a
gap at $\omega'_s = \omega_s + (v_s/2u_s)$.

\begin{figure}[h!]
\begin{center}
 \includegraphics[width=0.45\textwidth]{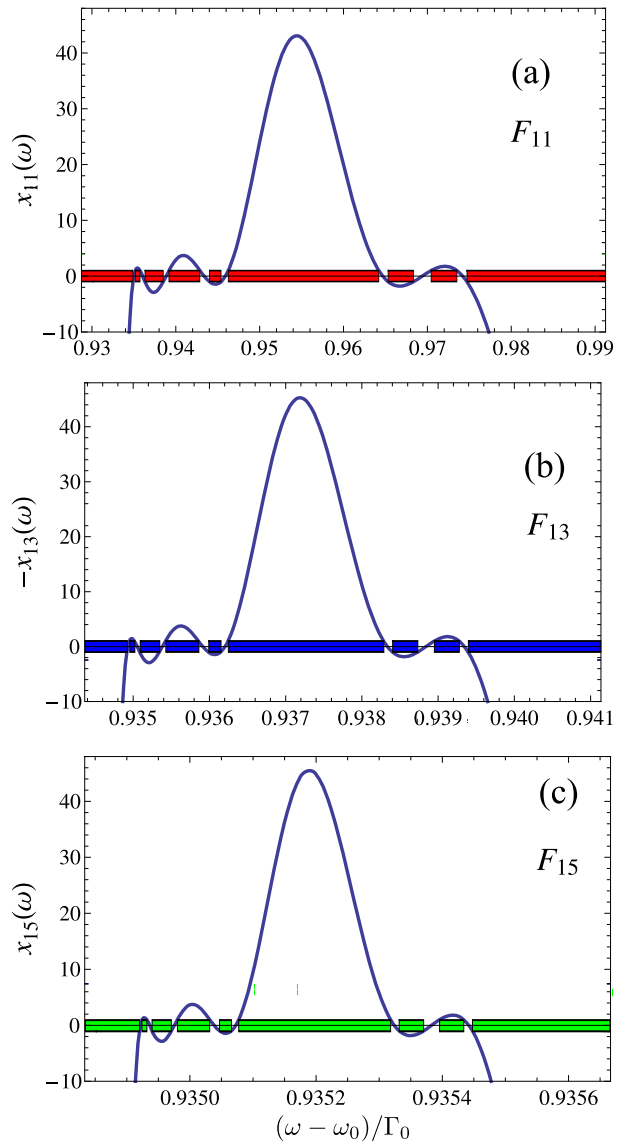}
\end{center}
 \caption{(Color online) Trace scaling  for the Fibonacci structures. The panels (a), (b) and (c) show the half-trace $x_j(\omega)$ for $j=11,13,15$, respectively. The filled ribbons indicate the regions of polariton band gaps, where $|x_j(\omega)|>1$.    Calculated for $\Gamma=0$
and other parameters same as in Fig.~\ref{fig:1}
  } \label{fig:5p}
\end{figure}

The second mechanism of gap formation is related to localized
exciton-polariton states rather than to the Fabry-Perot
interference and can be treated in terms of self-similarity
effects. Particularly, in the frequency range $|\omega - \omega_0|
\sim \Gamma_0$  the number of stop-bands increases while their
widths tend to zero as $N \to \infty$. As a result, the sequence
of the allowed and forbidden bands becomes quite intricate, see
Figs.~\ref{fig:5}(b)--\ref{fig:5}(d), and  locally resembles the
Cantor set.\cite{Kohmoto1984physlett} The most striking result in
Fig.~\ref{fig:5}(b) is similarity of the band structure of the
approximants with $j=11$ and $j=13$. On the other hand, the
spectrum for $j=13$ has a lot of narrow band gaps not resolved in
the scale of Fig.~\ref{fig:5}(b). Figures~\ref{fig:5}(c) and
\ref{fig:5}(d) present the same spectra in larger scales near the
right edge of the middle band gap, with the scale for $j=13$ being
$\lambda_+\approx  8$ times larger than that for
$j=13$. Matching the bandgap positions we prove an existence of
the spectral scaling in the Fibonacci QW structures. The scaling
index $\lambda_+$ specifies the ratio of the widths  of spectral
features of the structures with the order differing  by two. The
scaling properties hold not only for the band positions but for
the whole curves $x_j(\omega)$, as Fig.~\ref{fig:5p} demonstrates
for $j=11, 13$ and 15. One can see that the curves plotted in the
proper scales repeat each other almost exactly.

\begin{figure}[h!]
\includegraphics[width=0.49\textwidth]{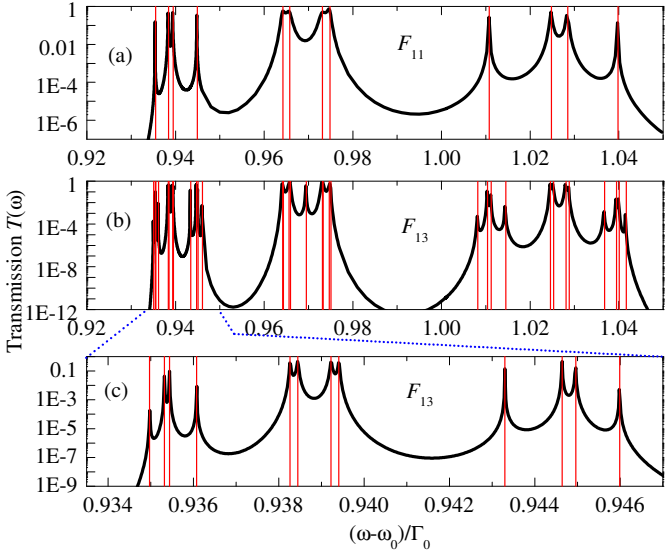}
 \caption{(Color online) Transmission spectra of the Fibonacci
quantum-well sequences of the order $j=$11(a), and   $j=$13 (b,c), containing $89$ and $233$ QWs, respectively.
 Vertical lines indicate the real parts of complex eigenfrequencies of the structures.
 Calculated for $\Gamma=0$
and other parameters same as in Fig.~\ref{fig:1}
} \label{fig:6}
\end{figure}

Now we turn from  the transfer matrix traces analysis, to the
complex eigenfrequencies \eqref{eq:oscillators} and to the
transmission spectra $T_j(\omega)$. The spectra $T_j(\omega)$ for
the resonant Fibonacci structures are shown in Fig.~\ref{fig:6}
for $j=11$~($N=89$ QWs) and  $j=13$ ($N=234$ QWs) in the frequency
region adjacent to the right edge of the middle band gap. The real
parts of the complex eigenfrequencies are indicated by the
vertical lines. The abscissa scales on Figs.~\ref{fig:6}(a) and
\ref{fig:6}(c) are the same as in Figs.~\ref{fig:5}(d) and
\ref{fig:5}(c). Examining Fig.~\ref{fig:6} we conclude, that the
scaling properties revealed by  Figs.~\ref{fig:5} and
\ref{fig:5p}, are also manifested in the optical spectra. Indeed,
comparing Fig.~\ref{fig:6}(a) and Fig.~\ref{fig:6}(b) we find that
the general shape of the spectra remains similar although  more
details appear when $N$ grows. On the other hand,  the relative
distances between the transmission peaks for $j=13$ in the large
scale agree with those  for $j=11$ in a smaller scale, cf.
Fig.~\ref{fig:6}(a) and Fig.~\ref{fig:6}(c). The positions of the
real parts of complex eigenfrequencies correspond to  the peaks in
the transmission spectra and exhibit the same scaling behaviour.
Such behavior is also demonstrated at the left edge of the middle
bandgap $\omega-\omega_0\sim -0.37\Gamma_0$, it is characterized
by the scaling index $\lambda_-\sim 16$.

The self-similarity of band structure of Fibonacci sequences with
the orders $j$ differing by 2 can be related to the
so-called ``band-edge''  cycle of the trace map
\eqref{eq:req1}.\cite{kohmoto1987}  This can be done by the
following consideration. If the half-traces $x_j$ for three
successive orders $j = j_0, j_0 + 1, j_0 +2$ are interrelated by
\begin{equation} \label{eq:fixpoint} x_{j_0 + 1} = -  \frac{x_{j_0}}{2 x_{j_0} - 1}\:,\:
x_{j_0 + 2} = -x_{j_0}
\end{equation}
then according to the recurrent equations \eqref{eq:req1}  the
values of $x_j$ for $j \geq j_0$ form a periodic sequence
\begin{align}\label{eq:cycle}
  x^*,-\frac{x^*}{2x^*-1}, -x^*,
\frac{x^*}{2x^*-1},-x^*,-\frac{x^*}{2x^*-1}, x^*,\ldots
\end{align}
where $x^* = x_{j_0}$.
Although the sequence \eqref{eq:cycle} repeats itself only after 6
iterations, the length of the cycle for  the absolute value of the trace is two.

 Considering $x_j$ as functions of the
frequency $\omega$ we introduce solutions $\omega^{(a)}(j_0)$ and
$\omega^{(b)}(j_0)$ of the first and second
equations~(\ref{eq:fixpoint}). The numerical calculation shows
that, for each $j_0$, in the vicinity of $\omega_0$ there exist
two pairs of solutions, $\omega^{(a)}_{\pm}(j_0)$,
and $\omega_{\pm}^{(b)}(j_0)$. Moreover, values
of $\omega^{(a)}_+(j_0)$ and $\omega_+^{(b)}(j_0)$ or
$\omega^{(a)}_-(j_0)$ and $\omega_-^{(b)}(j_0)$ merge with
increasing $j_0$, and one can introduce the asymptotic frequencies
$\omega^*_{\pm} = \omega^{(a)}_\pm (j_0 \to \infty) = \omega^{(b)}_\pm(j_0
\to \infty)$, with the values $\omega^*_+\approx 0.94\Gamma_0$ and
 $\omega^*_-\approx -0.37\Gamma_0$\:.

Our analysis made for  Fibonacci structures with different pair
values of $(h,h')$  shows that they also demonstrate analogous
scaling behaviour in the vicinity of $\omega_0$. The distance
$|\omega_+^*-\omega_-^*|$ between the scaling frequencies decreases
with the growth of the  barrier thicknesses. The  scaling indices increase, when the middle band gap becomes narrower. We have
established that there exists the following equation relating the
scaling indices and the  frequencies $\omega^*_{\pm}$:
\begin{equation}\label{eq:lambda2}
  \lambda_\pm \approx \Lambda_1+\Lambda_2\sqrt{I(\omega_\pm^*)}\equiv
\Lambda_1+\frac{\Lambda_2\Gamma_0}{|\omega_0-\omega_\pm^*|}
  \bigl|\sin\bigl[q(\omega_0)b(\tau-1)\bigr]\bigr|\:.
\end{equation}
The coefficients $\Lambda_1$ and $\Lambda_2$ are found to be close to 3 and 4,
respectively, and are independent of the structure parameters.
Since the value of $|\omega^*_+ - \omega_0|/ \Gamma_0 \approx 0.94$ exceeds
$|\omega^*_- - \omega_0|/ \Gamma_0 \approx 0.37$ the scaling coefficient
$\lambda_+ \approx 8$ is smaller than  $\lambda_- \approx 16$.

We have also analyzed   the
spatial structure of the excitonic polarization $P_m$ of the
eigenstates satisfying Eq.~\eqref{eq:oscillators}. In particular,
this distribution is characterized by the participation ratio $p =
\sum_{m=1}^N|P_m|^4/\bigl(\sum_{m=1}^N|P_m|^2\bigr)^2$, where the
sum runs over QW-lattice sites.\cite{Brezini1992} The parameter $p$
is a measure of the state localization-delocalization: for a
completely delocalized state $p \approx 1/N$, whereas $p \approx
1$ for a state tied to a single site. In the periodic Bragg
structure with $\bar{d} = \lambda(\omega_0)/2$, the superradiant
mode is described by the eigenvector $P_m = (-1)^m P_0$ and the
participation ratio $p=1/N$. In the Fibonacci QW chain the
participation ratio remains small for the superradiant mode as
well as for the subradiant modes described by the two-wave
approximation \eqref{eq:allfreqs}.
On the other hand, there exist localized states with large values of $p$ as well
as intermediate states.
 The
eigenstates with strongly localized character belong to the region
$|\omega - \omega_0|$ covering few $\Gamma_0$. The excitonic
polarization $P_m$ of such states is concentrated on a small
fraction of the QW chain and has a complex self-similar
structure.\cite{kohmoto1987}

Up to now we have limited ourselves only to the scaling features in the
region
$|\omega-\omega_0|\sim \Gamma_0$. At very high values of $N\gtrsim
1000$ the optical spectra become intricate even at
$|\omega-\omega_0|\gg \Gamma_0$. However, in the stopband regions
$(\omega_{\rm out}^{-}, \omega_{\rm in}^-)$ and $(\omega_{\rm
in}^+, \omega_{\rm out}^+)$ the reflection coefficient
$R_N(\omega) = |r_N(\omega)|^2$ remains close to unity. The
spectral pattern within the interval $(\omega_{\rm in}^-,
\omega_{\rm in}^+)$  strongly depends on values of $\Gamma$ and
$N$. For any nonradiative damping rate $\Gamma$ there exists a
finite value of the number of wells, $N(\Gamma)$, which separates
the structures into two categories. For $N > N(\Gamma)$, the
reflection spectrum is independent of $N$, $R_N(\omega;\Gamma)
\approx R_{\infty}(\omega;\Gamma)$,  and determined by the
exciton-polaritons localized within the area $0 < z <
z_{N(\Gamma)}$ and sensitive to the initial light. For QW
structures with $N < N(\Gamma)$, the light wave reaches the back
edge of the structure, reflects from this edge, propagates back
and participates in the Fabry-Perot interference resulting in the
oscillating reflectivity. This regime, $N < N(\Gamma)$, is well
described by the two-wave approximation except for the narrow
region near the exciton resonance frequency where the condition
(\ref{crit2}) is not satisfied. With decreasing $\Gamma$ the
critical number $N(\Gamma)$ infinitely increases while the
spectrum $R_{\infty}(\omega;\Gamma)$ continuously varies as
$\Gamma \to + 0$ and shows no saturation behaviour.

\subsection{Transmission spectra of the Thue-Morse quantum well structures}
We now turn to the Thue-Morse structures. The polariton
band-structure calculations performed for this system  lead to qualitatively
similar conclusions: two-wave
band gaps are already formed for small $j$, a middle narrow band
gap is always present, and a complicated sequence of the allowed
and forbidden bands arises around $\omega_0$. However, the
Thue-Morse structures have a very interesting specific properties,
most brightly manifested in their transmission spectra.
\begin{figure}[h!]\includegraphics[width=0.49\textwidth]{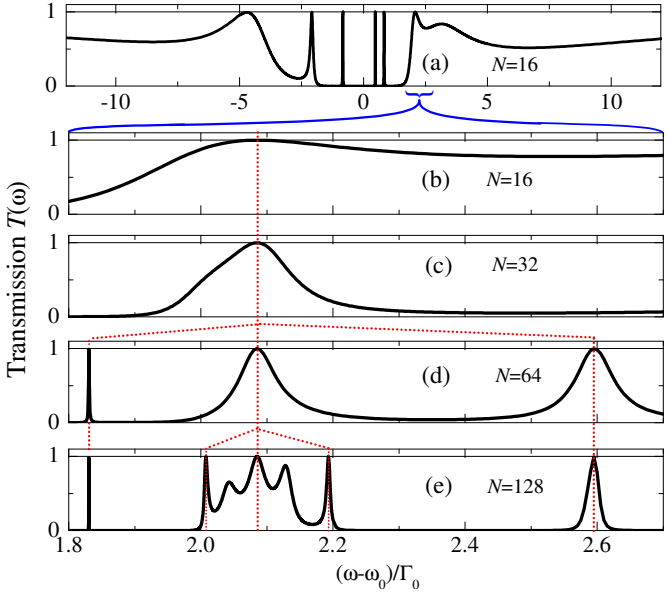}
 \caption{(Color online) Fine structure of the transmission
spectra calculated for Thue-Morse quantum-well sequences of order $j=4,5,6,7$ ($N=16\ldots 128$).
Calculated for $\Gamma=0$ and other parameters the same as in
Fig.~\ref{fig:1}.}\label{fig:7}
\end{figure}	

The transmission spectra are presented in Fig.~\ref{fig:7} for
different orders $j$ changing from $4$ ($N$ = 16 QWs) to $7$
($N=128$).
The spectrum has a complex structure with narrow peaks even for
$N=16$, see Fig.~\ref{fig:7}(a).
Figures~\ref{fig:7}(b)--\ref{fig:7}(d) show evolution of the
spectra with increasing the number of QWs.  The trace map \eqref{eq:req1b}
alone is not sufficient to obtain transmission spectra. Therefore, to analyze
the spectra
we use the standard properties of the trace map \eqref{eq:req1b}
and the antitrace maps \cite{Wang2000}
\begin{align}\label{eq:req2b}
 y_{j+1}&=2x_{j-1}[(2x_j-1)y_{j-1}+\tilde y_{j-1}]\:,\\ \nonumber
 \tilde y_{j+1}&=2x_{j-1}[(2x_j-1)\tilde y_{j-1}+ y_{j-1}]\:.
\end{align}
Here the half-antitrace $\tilde y_{j}$ corresponds
to the structure $\MT_j$ obtained from $\TM_j$ by the barrier
interchange $\A \leftrightarrow \B$, e.g., $\TM_2=\A\B\B\A$ and
$\MT_2=\B\A\A\B$. Contrary to the Fibonacci case, the trace map
\eqref{eq:req1b} for the Thue-Morse structures  has no  cycles of
type \eqref{eq:cycle}. Instead of such cycles,
Eqs.~\eqref{eq:req1b} and \eqref{eq:req2b} have the following
property \cite{Ryu1993}
\begin{equation}\label{eq:zero}
 \text{if $x_{j-2}(\omega)=0$ then}
 \begin{cases}
x_{j}(\omega)=x_{j+1}(\omega)=\ldots=1\:,\\
y_{j}(\omega)=y_{j+1}(\omega)=\ldots=0\:.
 \end{cases}
\end{equation}
As a consequence, the structure becomes transparent at this
particular frequency: $T_{j}(\omega)= T_{j+1}(\omega)=\ldots=1$.
The positions of the spectral features follow then from
Eq.~\eqref{eq:zero}. For example, the edges of the inner band gap
for the Thue-Morse structure, $\omega_L = \omega_0 - 0.83\Gamma_0$
and $\omega_R = \omega_0 + 0.47 \Gamma_0$, can be found from the
equations $x_1(\omega_L)=0$ and $x_2(\omega_R)=0$, respectively.

Whenever the half-trace $x_{j-2}(\omega)$ vanishes at some
frequency $\omega = \omega_1$ and therefore, $x_j(\omega_1)= 1$,
there exist two neighboring frequencies, $\omega_2 < \omega_1$ and
$\omega_3 > \omega_1$, such as $x_j(\omega_{2,3}) = 0$.
As a consequence, $x_{j+2}(\omega_{2,3}) = 1$ and
the transmission coefficient for the structure $\TM_{j+2}$ at these frequencies
reaches unity. Thus the number of transmission peaks increases with
the growth of the structure order: each unitary peak in the spectrum of the
structure $\TM_j$ (i) persists in the spectra of the structures
$\TM_{j+1},\TM_{j+2},\ldots $ of higher orders and (ii) leads to  appearance
of two more adjacent peaks for the structure $\TM_{j+2}$.  An example
of such ``tree'' of trifucations is indicated in
Figs.~\ref{fig:7}(b)--\ref{fig:7}(e) by dashed lines. The single transmission
peak of the structure $\TM_4$ ($N=16$) at $\omega\approx \omega_0+2.1\Gamma_0$
 has  generated the two  peaks at
$\omega\approx \omega_0+1.8\Gamma_0$ and
$\omega\approx \omega_0+2.6\Gamma_0$ for
the structure $\TM_6$ ($N=64$), cf. Fig.~\ref{fig:7}(b) and
Fig.~\ref{fig:7}(d). The characteristic
widths of the spectral features tend to zero proportional to a power the
structure length $N_j^{- \alpha} = 2^{- j \alpha}$,
where $\alpha$ is a frequency-dependent positive index.
Transmission spectra can also have non-unity maxima, see
Fig.~\ref{fig:7}(e). These peaks do not correspond to any special
values of $x_j$ and $y_j$ and their positions depend on $j$.  The spatial
distribution of the electric field on the frequencies with unitary transmission
has a so-called ``lattice-like'' shape, specific for the Thue-Morse
structures.\cite{Ryu1992}

The spectra presented in Fig.~\ref{fig:7} are calculated for the Bragg
structure with $q(\omega_0)(a+b)=2\pi$. An interesting property
of the Thue-Morse lattices, satisfying the Bragg condition
$q(\omega_0)(a+b)=\pi,2\pi\ldots$, is the mirror symmetry between
the transmission spectra $T(\omega)$ and $\widetilde{T}(\omega)$,
of the structures $\TM_j$ and $\MT_{j}$, holding in the region
$|\omega-\omega_0|\ll\sqrt{\Gamma_0\omega_0}$ in which case $q =
q(\omega)$ in Eq.~(\ref{eq:MD}) can be approximated by
$q(\omega_0)$. The spectra are symmetrical with respect to
$\omega=\omega_0$: $\widetilde{T}(\omega)=T(2\omega_0-\omega)$.

\section{Conclusions}\label{sec:conl}
We have developed a theory of exciton-polaritons in
photonic-quasicrystalline and aperiodic MQW structures having two
different values of the interwell distances, $\A$ and $\B$. The
approach based on the two-wave approximation has been extended to
derive both the dispersion equation and analytical formulas for the
reflectance and transmittance spectra from aperiodic MQW
sequences. This  approximation  successfully describes the
pattern of the optical spectra including the pair of stop-bands
and interference fringes between them but stops being valid in a
spectral range $|\omega - \omega_0| \sim \Gamma_0$ near the
exciton resonant frequency $\omega_0$. In the Fibonacci QW chains
with small values of the exciton nonradiative damping rate,
$\Gamma < \Gamma_0$, the transmittance spectra and polariton band
structure reveal, at two edges of the narrow inner band gap, a
complicated and rich structure demonstrating scaling invariance
and self-similarity. It has been shown that this structure can be
related to the ``band-edge'' cycle of the trace map.  The fine
structure of optical spectra in the Thue-Morse MQWs can be
interpreted in terms of zero reflection frequencies, or
frequencies of unity transmittance.\cite{Ryu1993}

\acknowledgements {This work was supported by RFBR and the
``Dynasty'' Foundation -- ICFPM. The authors thank G.~Khitrova and
H.M.~Gibbs for helpful discussions.}

\setcounter{equation}{0}
\renewcommand{\theequation}{A\arabic{equation}}
\section*{Appendix} \label{sec:Appendix}

Here we will analyze the second order of the perturbation theory
at the point $K = - K' = G_{h^{\vphantom{'}}h'}/2 \equiv q_0$ and
confirm the high accuracy of the stop-band edges defined by
Eqs.~(\ref{eq:ws}).

We consider the terms in Eqs.~(\ref{eq:fourier}) proportional to
the structure-factor coefficients $f_{h-g, h' - g'}^*$ as a
perturbation. Then keeping the second order contributions we
obtain the following equations for the amplitudes $E_K$ and
$E_{K'} = E_{-K}$:
\begin{gather}\label{eq:2ndOrder}
[q^2 - K^2 + \chi ( 1 + \chi \bar{d}^2 \zeta_{11})] E_K +\chi (f^*_{hh'} + \chi \bar{d}^2
\zeta_{12}) E_{-K}  = 0\:, 
\nonumber
\\\chi ( f_{hh'}  + \chi \bar{d}^2 \zeta_{21})  E_{K}
+ [q^2 - K^2 + \chi ( 1 + \chi \bar{d}^2 \zeta_{22})] E_{-K} = 0
\:. 
\end{gather}
Here $\chi = 2q \xi = 2q \Gamma_0/[\bar{d}(\omega_0 - \omega +
{\rm i} \Gamma)]$,
\begin{eqnarray}\label{eq:zeta11}
&&\zeta_{11} =\zeta_{22} = \frac{1}{\bar{d}^2}\!\sum\limits_{\substack{(g,g') \neq
(h,h')\\(g,g') \neq (0,0)}} \frac{\vert f_{gg'}
\vert^2}{(K - G_{gg'})^2 - K^2}\:,\\
\label{eq:zeta21}
&&\zeta_{21} =\zeta_{12}^*= \frac{1}{\bar{d}^2}\!\sum\limits_{\substack{(g,g')
\neq (h,h')\\(g,g') \neq (0,0)}} \frac{f_{h-g,h'-g'} f_{gg'}
}{(K - G_{gg'})^2 - K^2}\:.
\end{eqnarray}
We remind that, for the Fibonacci chains, the structure-factor
coefficients are given by
\begin{equation}
 f_{gg'} = \frac{\sin S_{gg'}}{S_{gg'}} \exp\Bigl({\rm i}
\frac{\tau-2}{\tau} S_{gg'}\Bigr),\quad S_{gg'} = \frac{\pi
\tau}{\tau^2+1}(\tau g'-g)\:.\label{eq:ffc2}
\end{equation}

The applied perturbation theory differs from the standard one by
the presence of terms in the corresponding sums with the
denominators $(K - G_{gg'})^2 - K^2$ arbitrarily close to zero.
However the sums are finite because the smallness of the
denominator at particular values of $g$ and $g'$ is compensated by
much smaller values of the numerators for the same values of $g,
g'$. The convergence of the sum (\ref{eq:zeta11}) for pairs $g,
g'$ where $G_{gg'} \approx 0$ or $G_{gg'} \approx G_{hh'}$  is
checked as follows. Taking into account the symmetry property
$|f_{gg'}| = |f_{-g,-g'}|$ we can perform the following
replacement in Eq.~(\ref{eq:zeta11})
\[
\frac{\vert f_{gg'} \vert^2}{(K - G_{gg'})^2 - K^2} \to
\frac{\vert f_{gg'} \vert^2}{G_{gg'}^2 - G^2_{h h'}}\:.
\]
Therefore, this sum converges for the pairs $g, g'$ with $G_{gg'}$
tending to zero. Now let us consider the sequence of pairs $(g,g')
= (h + k, h' + k')$, where
\[
k = m F_j, k' = - m F_{j+1}\:,
\]
$F_j$ are the Fibonacci numbers and $m$ is any integer $\pm 1, \pm
2...$ different from 0. Taking into account that
\begin{gather}\label{ident} F_j = \frac{\tau^j - (1 - \tau)^j}{2
\tau - 1} \:,\:\frac{F_j}{\tau} - F_{j-1} = \frac{(1 -
\tau)^{j-1}}{\tau}\nonumber
\end{gather}
the chosen sequence possesses the properties
\begin{gather*}
G_{gg'} - G_{hh'} = - m \frac{2 \pi }{\bar{d}} \frac{(1 - \tau)^{j-1}}{\tau}\:,\\
S^2_{gg'} (j \to \infty) \approx \left( \frac{\pi m \tau}{2 \tau - 1} \right)^2 \tau^{2(j-1)}\ .
\end{gather*}
and, hence, with increasing $j$ one has
\[
|G_{gg'} - G_{hh'}| S^2_{gg'} \propto |m|^3 \tau^{j}
\]
which means that the sum
\[
\sum\limits_{m,j}|G_{gg'} - G_{hh'}|^{-1} S^{-2}_{gg'}
\]
converges. The convergence in Eq.~(\ref{eq:zeta21}) is checked in
a similar way.

Numerical calculation performed for the Fibonacci QW chain with
$(h,h')=(1,0)$ shows that $|\zeta_{11}|$ and $|\zeta_{21}|$ are
both smaller than $0.1$. The prefactor $\chi \bar{d}^2$ in Eq.
\eqref{eq:2ndOrder} is small as compared with $f_{hh'}$ whenever
\begin{equation}\label{eq:far}
{\rm max} \{ |\omega_0 - \omega| , \Gamma \} \gg \frac{2 \pi \Gamma_0}{|f_{hh'}|} \left( h + \frac{h'}{\tau} \right)\:.
\end{equation}
We conclude that, for the wave (\ref{eq:bloch}) with $K = q_0$,
the contributions from the wavevectors $G_{gg'}$ different from 0
and $G_{hh'}$ are negligible within the applicability range of Eq.
\eqref{eq:far}.



\end{document}